\documentclass[nofootinbib,aps,prd,10pt,superscriptaddress,showpacs,preprintnumbers,eqsecnum]{revtex4-1}

\usepackage{amsmath}
\usepackage{dsfont}
\usepackage{multirow}
\usepackage{graphicx}
\usepackage{mathrsfs}
\usepackage{amsfonts}
\usepackage{amssymb}
\usepackage[dvips]{color}

\begin{document}
\newcommand\be{\begin{equation}}
\newcommand\ee{\end{equation}}
\newcommand\bea{\begin{eqnarray}}
\newcommand\eea{\end{eqnarray}}
\newcommand\bseq{\begin{subequations}} %solo con amsmath
\newcommand\eseq{\end{subequations}}
\newcommand\bcas{\begin{cases}}
\newcommand\ecas{\end{cases}}
\newcommand{\p}{\partial}
\newcommand{\f}{\frac}

\title{Cosmological-Billiards Groups and self-adjoint BKL Transfer Operators}

\author{Orchidea Maria Lecian}
\email{lecian@icra.it}
\affiliation{Sapienza University of Rome, Physics Department and ICRA, Piazzale Aldo Moro, 5 - 00185 Rome, Italy}

%\today

\begin{abstract}
The Selberg trace formula for cosmological billiards is assessed. The Selberg trace formula for cosmological billiards is made descend from the properties of the conjugacy classes of the congruence subgroup of the desymmetrised modular group which constitutes the cosmological-billiard group, and is derived from the qualities of the corresponding Bogomolny semiclassical transfer operator, of which the qunatum implementation is upgraded.\\ 
Cosmological billiards arise as a map of the solution of the Einstein equations, when the most general symmetry for the metric tensor is hypothesized, and points are considered as spatially decoupled in the asymptotic limit towards the cosmological singularity, according to the BKL (Belinski Khalatnikov Lifshitz) paradigm. In $4=3+1$ dimensions, two kinds of cosmological billiards are considered: the so-called 'big billiard' which accounts for pure gravity, and the 'small billiard', which is a symmetry-reduced version of the previous one, and is obtained when the 'symmetry walls' are considered.\\
The solution of Einstein field equations is this way mapped to the (discrete) Poincar\'e map of a billiard ball on the sides of a triangular billiard table, in the Upper Poincar\'e Half Plane (UPHP).\\
The billiard modular group is the scheme within which the dynamics of classical chaotic systems on surfaces of constant negative curvature is analyzed.
The  periodic orbits of the two kinds of billiards are classified, according to the different symmetry-quotienting mechanisms.\\
The differences with the description implied by the billiard modular group are investigated and outlined.\\
The comparison is duely done with the full symmetry-unquotiented system.\\
In the quantum regime, the eigenvalues (i.e. the sign that wavefunctions acquire according to quantum BKL maps) for periodic phenomena of the BKL maps on the Maass wavefunctions are classified.
The complete spectrum of the semiclassical operators which act as BKL map for periodic orbits is obtained.\\
Differently form the case of the modular group, here it is shown that the semiclassical transfer operator for Cosmological Billiards is not only the adjoint operator of the one acting on the Maass waveforms, but that the two operators are the same self-adjoint operator, thus outlining a different approach to the Langlands Jaquet correspondance.

\end{abstract}


\maketitle
\section{Introduction\label{section1}}
The solution to the Einstein field equations in the asymptotic limit towards the cosmological singularity corresponds, within the BKL (named after Belinski, Khalatnikov and Lifshitz) Ref. 1, Ref. 2, Ref.3, Ref. 4, Ref.6 Ref. 7 paradigm, for which space gradients are neglected with respect to time derivatives, under the most general assumptions for the symmetries of the metric tensor, to the asymptotic limit towards the cosmological singularity of a Bianchi IX Universe. This way, the Einstein field equations for a Generic Universe correspond to a system of ordinary differential equations, whose symmetry for homogeneous universes and for inhomogeneous universe, $\Gamma_2(PGL(2,C))$ and c, respectively, allow one to define the corresponding billiard systems on the Upper Poincar\'e Half Plane, and the corresponding billiard maps for the discretized dynamics Ref.11, Ref.12, Ref. 13.\\
\\
The conjugacy subclasses of the Billiard Modular Group define the $\Gamma_2(PGL(2,C))$ congruence subgroup of $PGL(2,C)$. The discretization of the dynamics allows for the definition of language codes, i.e. the composition of transformations that describe the continuous billiard dynamics and according to the time evolution of the Einstein field equations.\\
\\
The analysis is brought according to the prescriptions of \cite{belinskibook} Chapter 4, and, in particular, is implemented after the tools developed in \cite{Lecian:2013cxa}.\\
\\
The unquotiented cosmological billiard groups therefore describe the (time) evolution of the dynamics, for which the insertion of any composition of transformations corresponding to the Identity within the unquotiented maps would correspond to (with respect to the time evolution and therefore with respect to the symmetries of the solution of the Einstein field equations) unphysical sequences of trajectories.\\
\\
The eigenvalues for the BKL quantum operators that constitute the quantum maps and the semiclassical ones are defined according to the language codes of the big billiard and of the small billiard.\\ 
Cyclic identities for periodic orbits of the big billiard map and of the small billiard map define the parity of the (with respect to the corresponding WDW equation) suitable Maass wavefunctions and therefore exactly solve the Selberg trace formula for cosmological billiards.\\
\\
The paper is organized as follows.\\
In Section \ref{section2}, the Billiard Modular Group is defined.\\
In Section \ref{section3}, the conjugacy subclasses and the definition of periodic orbits of the Big Billiard Group are defined.\\ 
In Section \ref{section4}, the Small Billiard group is implemented.\\
In Section \ref{section4a}, the analysis of dynamics of the complete system is recalled.\\
In Section \ref{section5a}, the full (symmetry-unquotiented) system is analysed within the tools here developped.\\
In Section \ref{section5}, a comparison of the systems is outlined.\\
I Section \ref{secmath}, th mathematical tools propedeutical of the definition of the Selberg trace formula of cosmologcal billiards are recalled.\\
In Section \ref{section6}, the Selberg trace formula for comsmological billiards is assessed: the quantum regime and the semiclassical transition are analyzed by the definition of the semiclassical Poincar\'e surface of section for cosmological billiards and the implementation of the Selberg trace formula according to the sign acquired by the eigenvalues of the quantum BKL maps for periodic Cosmological Billiards orbits.\\
Outlook and Perspectives are briefly stated in Section \ref{section7}.\\
Brief concluding remarks follow in Section \ref{section8}.\\
The Appendix \ref{appendixa} is devoted to the analysis of the symmetry-quotiented maps, for the comparison with the recovery of the BKL quantum numbers after the implementation of the prescriptions of \cite{belinskibook} Chapter 4 with the tools developed in \cite{Lecian:2013cxa}.

\section{The Billiard Modular Group\label{section2}}
\begin{figure*}[htbp]
\begin{center}
\includegraphics[width=0.7\textwidth]{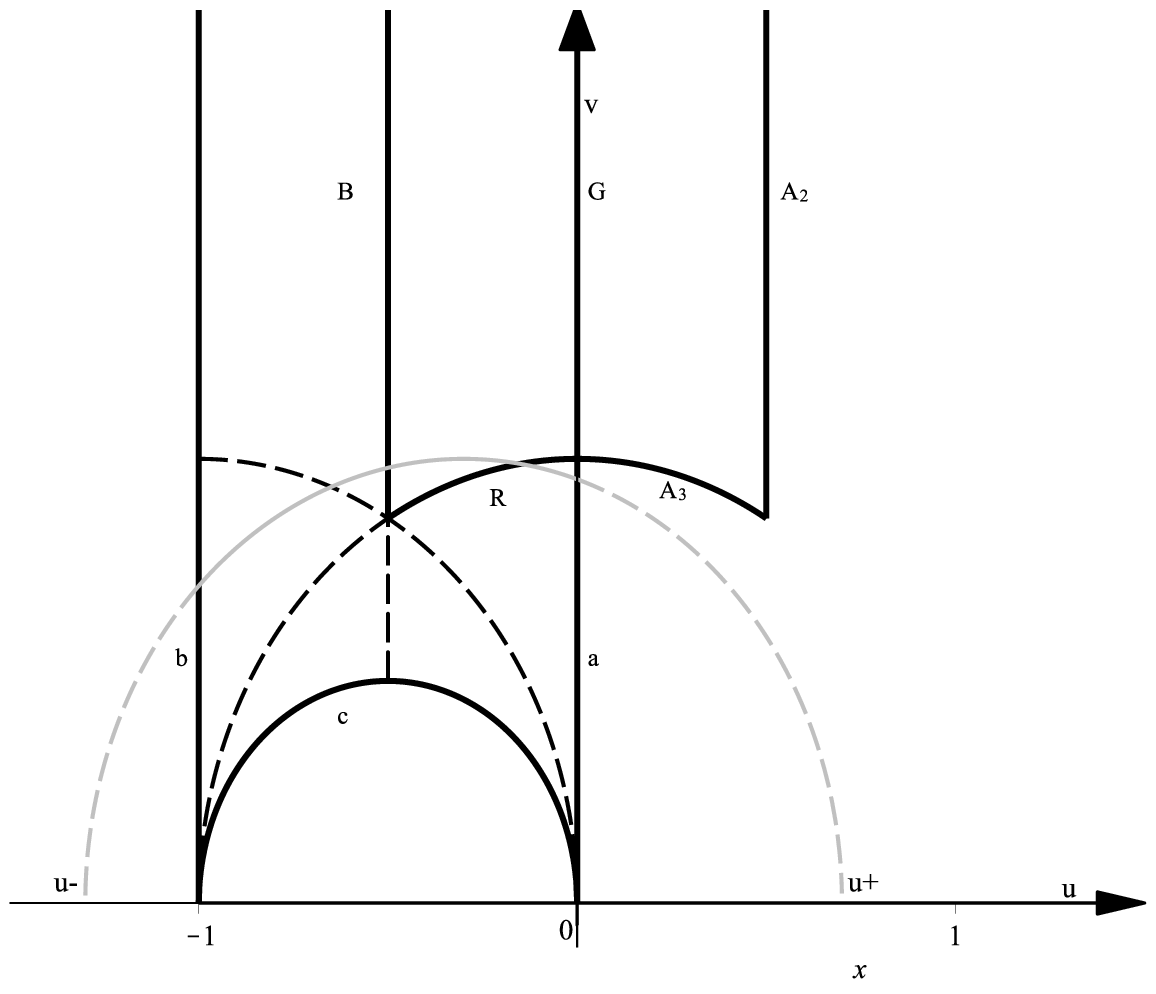}
\caption{\label{figura1res} The domains of the big billiard $\Gamma_2(PGL(2,C))$, the small billiard, the $SL(2,Z)$ group, the $SL(2,C)$ group and the billiard modular group. In particular, the domain of the big billiard is delimited by the geodesics $u=0$, $u=-1$, $u^2+u=0$; the domain of the small billiard, which coincides with that of the $SL(2,Z)$ group, is delimited by the geodesics $u=0$, $u=-1/2$ and $u^2+v^2=1$; the domain of the $SL(2,C)$ group is delimited by the geodesics $u=1/2$, $u=-1/2$ and $u^2+v^2=1$; the domain of the billiard modular group is delimited by the geodesics $u=0$, $u=1/2$ and $u^2+v^2=1$. They are all plotted by solid black lines. The simmetry lines of the big billiard are represented by the dashed black lines. An oriented geodesic is drawn as gray dashed circle, and the oriented endpoints $u^+$ and $u^-$ are indicated on the $u$ axes.}
\end{center}
\end{figure*}

The Billiard Modular Group (BMG) is defined on the asymmetric domain delimited by the geodesics $\mathcal{A}$, $\mathcal{B}$, $\mathcal{C}$, such that 
\begin{subequations}\label{bmgdomain}
\begin{align}
&\mathcal{A}: u=0,\\
&\mathcal{B}: u=\tfrac{1}{2}\\
&\mathcal{C}: u^2+v^2=1,
\end{align}
\end{subequations}
for which the following transformations are defined
\begin{subequations}\label{bmg}
\begin{align}
&\mathcal{A}z=-\bar{z},\label{matha}\\
&\mathcal{B}z=1-\bar{z},\label{mathb}\\
&\mathcal{C}z=\tfrac{1}{\bar{z}}\label{1z}.
\end{align}
\end{subequations}
The Modular Billiard group domain is depicted in Figure \ref{figura1res} for comparison with the Big Billiard domain.
\paragraph{The language code for the BMG}
According to these transformation for the asymmetric domain (\ref{bmgdomain}), any matrix of the BMG can be written as one of the following
\begin{itemize}
	\item $\mathcal{A}$, $\mathcal{B}$, $\mathcal{C}$, $\mathcal{AC}$, $\mathcal{BC}$, $(\mathcal{BC})^2$;
	\item $I$, $\mathcal{BA}\tt{T}^n$, $\mathcal{B}\tt{T}^n$, $\tt{T}^n\mathcal{A}$, $\mathcal{B}\tt{T}^n\mathcal{A}$, $\mathcal{BA}$;
	\item $M\mathcal{C}\tt{T}^{n_1}\mathcal{C}\tt{T}^{n_2}...\mathcal{C}\tt{T}^{n_k}N$,
\end{itemize}
where the matrix $I$ is the identity matrix, the matrix $\tt{T}$ is defined as $\tt{T}=\mathcal{AB}$, such that $\tt{T}z=z-1$, $\tt{T}^n$ is its $n$-th iteration, the matrix $M$ can be one of the following: $I$, $\tt{T}^n$, $\mathcal{B}\tt{T}^{n}$, $\mathcal{CB}$, $\mathcal{B}$; the matrix $N$ can be one of the following: $I$, $\mathcal{C}$, $\mathcal{A}$, $\mathcal{A}\tt{T}^{n}$, $\mathcal{CA}$. This way, the \textit{language code} for the BMG is defined.\\
By a suitable transformation (conjugation), any matrix of the BMG can be written as one of the following
\begin{itemize}
	\item elliptic transformations $\mathcal{A}$, $\mathcal{B}$, $\mathcal{C}$, $\mathcal{AC}$, $\mathcal{BC}$, $(\mathcal{BC})^2$;
	\item parabolic transformations $\tt{T}^n$;
	\item hyperbolic transformations $\mathcal{C}\tt{T}^{n_1}\mathcal{C}\tt{T}^{n_2}...\mathcal{C}\tt{T}^{n_k}$.
\end{itemize}
\paragraph{Action of the BMG on the oriented endpoints}
The periodic orbits of the BMG can be described by hyperbolic transformations by a sequence of integers $n_1, n_2, ..., n_k$, after identification of circular permutations, for which reduced matrices are defined.\\
A periodic orbit of the BMG is defined by two oriented endpoints of a geodesics, $-1<x'<0$ and $x>1$, which are invariant under the action of hyperbolic transformations, which define a quadratic equation with integer coefficients, i.e.
\begin{subequations}
\begin{align}
&\mathcal{C}\tt{T}^{n_1}\mathcal{C}\tt{T}^{n_2}...\mathcal{C}\tt{T}^{n_k}x'=x',\\
&\mathcal{C}\tt{T}^{n_1}\mathcal{C}\tt{T}^{n_2}...\mathcal{C}\tt{T}^{n_k}x=x,
\end{align}
\end{subequations}
such that the continued-fraction decomposition of $x$ reads
\begin{equation}
x=n_1+\tfrac{1}{n_2+\tfrac{1}{...n_k+\tfrac{1}{n_1+...}}},
\end{equation}
the continued-fraction decomposition of $x'$ reads
\begin{equation}
-x'=n_1+\tfrac{1}{n_2+\tfrac{1}{...n_k+x'}}.
\end{equation}
The code of matrices, the continued-fraction expansion of the oriented endpoints of the geodesics and the expression of the roots of the quadratic forms are equivalent for the definition of the conjugacy subclasses of the BMG, and can be used to gain information one from the others.
\subsection {Comparison with the SL(2,Z) group}
The group $SL(2,Z)$ is defined on the symmetric fundamental domain delimited by the sides $A_1$, $A_2$, $A_3$, described by
\begin{subequations}\label{sl2}
\begin{align}
&A_1: \ \ u=-\tfrac{1}{2},\\
&A_2: \ \ u=\tfrac{1}{2},\\
&A_3: \ \ u^2+v^2=1,\label{A3}
\end{align}
\end{subequations}
with the transformations
\begin{subequations}\label{sl}
\begin{align}
&T(z)=z+1,\\
&S(z)=-\tfrac{1}{z}\label{S}
\end{align}
\end{subequations}
The sides are identified as
\begin{subequations}
\begin{align}
&T: \ \ A_1\rightarrow A_2,\\
&S: \ \ A_3\rightarrow A_3.\label{S1}
\end{align}
\end{subequations}
In particular, it is straightforward verified that the transformation $S$ in Eq. (\ref{S}) acts in Eq. (\ref{S1}) by identifying the $u>0$ part of the side $A_3$ in Eq. (\ref{A3}) with the $u<0$ part of the same side, and viceversa.\\ 
\\
The asymmetric domain of the BMG (\ref{bmgdomain}) is obtained by a suitable desymmetrizzation of the symmetric domain of the group $SL(2,Z)$.\\
By comparison with the transformations that define the BMG in Eq. (\ref{bmg}), one learns that the action of the transformation $\mathcal{C}$ in Eq. (\ref{1z}) can be interpreted as a symmetry-quotienting mechanism induced on the desymmetrized domain (\ref{bmgdomain}) of the BMG. This symmetry-quotienting mechanism defines the language code of the BGM, such that periodic orbits of the BMG are classified according to this convention.\\ 
\\
%%%%%%%%%%%%%%%%%%%%%%%%%%%%%%%%%%%%%%%%%%%%%%%%%%%%%%%
\section{The big billiard group\label{section3}}
The big-billiard group (BBG) is obtained from the big billiard table, i.e. a domain defined by the three sides $a$, $b$, $c$,  
\begin{subequations}\label{bbz1000}
\begin{align}
&a: u=0\\
&b: u=-1\\
&c: u^2+u+v^2=0, 
\end{align}
\end{subequations}
for which bounces against the billiard sides are expressed by the following transformations on the UPHP
\begin{subequations}\label{bbz}
\begin{align}
&Az=-\bar{z},\\
&Bz=-\bar{z}-2,\\
&Cz=-\tfrac{\bar{z}}{2\bar{z}+1}.
\end{align}
\end{subequations}
The unquotiented big-billiard map $\mathcal{T}$ consists of a suitable composition of the transformations (\ref{bbz}); the sequence of this composition is obtained by the continued-fraction decomposition of the $u^+$ variable.\\
Periodic orbits of the big billiard are a phenomenon which is more complicated than its symmetry-quotiented versions \cite{Damour:2010sz}: there exists a particular Kasner transformation $k_*$ (which depends on $m$ and on the considered periodic orbit of $\mbox{\large{$\mathsf{T}$}}$) such that
\be\label{171}
{\mathcal T}^m (u^- , u^+) = k_* (u^- , u^+) \, .
\ee
The set of six Kasner transformations is a realization of the $S_3$ permutation group (of order $3!=6$). In fact, this permutation group consists of the identity, $3$ transpositions [$(12), (23)$ and $(31)$], and $2$ cyclic transformations [$(213)$ and $(321)$].We recall that the \textit{order} of a particular group element, such as $k_*$, is the smallest integer $p$ such that $k_*^p=k_0$. As a transposition is of order $2$, and a cyclic permutation, ($123$) or ($321$), of order $3$, we see that the order $p$ of $k_*$ must be equal to $p=1, 2$ or $3$. Therefore, by iterating (\ref{171}), we get
\be
{\mathcal T}^{mp} (u^- , u^+)=k_*^p (u^- , u^+)= (u^- , u^+) \, ,
\ee
and $mp$ will be the smallest such integer. In other words, $(u^- , u^+)$ is the initial point of a periodic orbit under the unquotiented billiard map ${\mathcal T}$, with period $pm$, where $p=1,2,3$ is the order of $k_*$. 

\subsection{Cyclic identities}
Given any $x, y$ in $A, B, C$, periodic orbits are defined by imposing the condition (as alternatively solved in \cite{lecian2013selb})
\be
\prod_{n_i}\it{T}_{(x, y; n_i)}z=z
\ee
where the epoch map $\it{T}_{(x, y; n_i)}$ for a $n_i$-epoch BKL era of the $xy$ type is defined as
\begin{itemize}
	\item $\it{T}_{(x, y; n_i)}=y(xy)^{\tfrac{n_i-1}{2}}$, $n_i$ odd,
	\item $\it{T}_{(x, y; n_i)}=y(xy)^{\tfrac{n_i-1}{2}}x$, $n_i$ even,
\end{itemize}
with $\sum_in_i=mp=n$.\\

%Cyclic identities for the Big Billiard Group language code are defined as a suitable solution of Eq. (\ref{171}):

\section{The small-billiard group\label{section4}}
The small billiard is delimited by the sides $G$, $B$, $R$, defined as 
\begin{subequations}\label{smdomain}
\begin{align}
&G: \ \ u=0,\\
&B: \ \ u=-\tfrac{1}{2},\\
&R: \ \ u^2+v^2=1.
\end{align}
\end{subequations}
The transformation that describe the bounces of the billiard ball against the sides of the small billiard table (\ref{smdomain}) are those that define the $SL(2,Z)$ group, i.e. 
\begin{subequations}\label{smalltrasf}
\begin{align}
&R_1(z)=-\bar{z},\\
&R_2(z)=-\bar{z}+1,\\
&R_3(z)=\tfrac{1}{\bar{z}},
\end{align}
\end{subequations}
where no identification among the sides is present, and no symmetry-quotienting mechanisms for the $R$ side of the small billiard table (\ref{smdomain}) is assumed.\\
The action of the transformations (\ref{smalltrasf}) on the oriented endpoints of the oriented geodesics that define the trajectories of the billiard ball on the small billiard table is obtained by imposing $v\equiv 0$ on the UPHP variable $z=u+iv$, and results in the small-billiard billiard map $t$, which acts diagonally on the reduced phase space variables $u\equiv u^\pm$ as  
\begin{subequations}\label{t}
\begin{align}
&Gu= -u,\\
&Bu= -u-1\\
&Ru= \tfrac{1}{u}.
\end{align}
\end{subequations}
Epochs on the small billiard table are defined as any trajectory joining any two walls of the small billiard sides; eras for the small billiard are defined as a succession of epochs starting from the side $R$.\\
%%%%%%%%%%%%%%%%%%%%

\section{The complete Billiard system\label{section4a}}
\begin{figure*}[htbp]
\begin{center}
\includegraphics[width=0.7\textwidth]{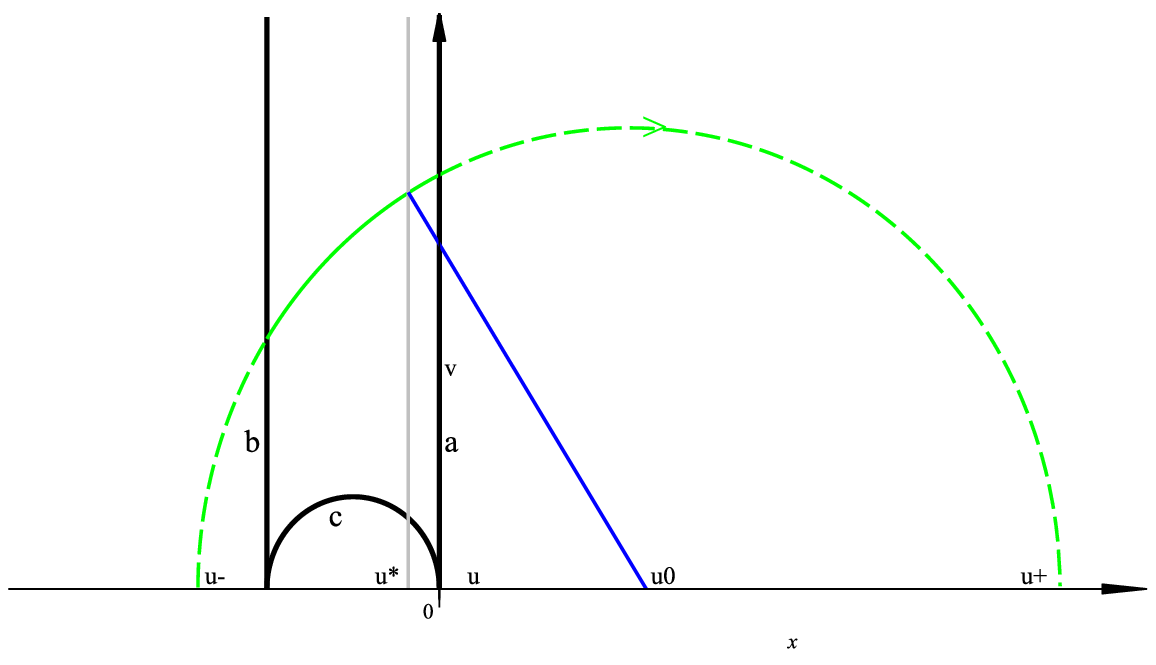}
\caption{\label{figuraustar} The angle $\theta$ is comprehended within the axis of abscissae and the radius $r$ of the geodesics connecting the center of the geodesics $u_0$ with the intersection point between the generalized Poincar\'e surface of section $u_*=const$ and the same geodesics, such that $\cos\theta = (u_0-u^*)/r$.}
\end{center}
\end{figure*}

It is possible to analyse the dynamics of the complete billiard system after the specifications of \cite{Lecian:2013cxa}.\\
The dynamics of the unquotiented Billiard system s described of a sequence f simple reflections. The conjugacy subclasses of this congruence subgroup of the desymmetrized modular group describe the dynamics of the Big-Billiard system, when the reflections associated to one of the operators in the Left Column of Table \ref{table1} are evaluated at a particular value of the $v$ variable, i.e. at $v=0$, the Big-Billiad unqutiented map of the $u$ variable is recovered.\\
The BKL map is traced is one applies the inverse of the Kasner operators in Table \ref{table1} to the unquotiented dynamics.\\
The BKL epoch map is therefore recovered after considering the reflections for the trajectories with $u>1$ on the UPHP, and by applying the pertinent inverse of the Kasner operator.\\
The BKL era maps is recovered, accordingly, by keeping track of the number and quality of reflections.\\
The advantage of defining a BKL epoch map and a BKL era map on the unquotiented system is the possibility to analyse the complete unquotiented phase space.\\
The dynamics of the unquotiented phase space is apt for establishing Poincar\'e return maps, as illustrated in Fig. \ref{figuraustar} on the UPHP, and, more importantly, allows one to define a surface of section in the complete phase space.\\
The particular choice of a Poincar\'e return map of the unquotiented phase space delineated on the UPHP at $v=0$ allows one to recover the BKL quantum numbers, when the quantum regime is implemented. 
\begin{table}
\begin{center}
%\tbl{The Kasner transformations.\newline
%}
   { \begin{tabular}{ || l |  l || l || l || }
   \hline
      $ca\rightarrow ba$ & $K_1$ & $z\rightarrow 1/\bar{z}$ & $ R_3$\\ \hline
      $ac\rightarrow ba$ & $K_2$ & $z\rightarrow -(1+z)/z$ & $(R_1R_2)(R_1R_3)$\\ \hline
      $bc\rightarrow ba$ & $K_3$ & $z\rightarrow -\bar{z}/(\bar{z}+1)$ & $(R_1R_3)(R_2R_1)R_3$\\ \hline
      $cb\rightarrow ba$ & $K_4$ & $z\rightarrow -1/(z+1)$ & $(R_3R_1)(R_2R_1)$ \\ \hline
      $ab\rightarrow ba$ & $K_5$ & $z\rightarrow -\bar{z}-1$ & $R_1R_2R_1$\\ \hline
      $ba\rightarrow ba$ & $K_0$ & $z\rightarrow z$ & $I$ \\ \hline
    \end{tabular}}
		\end{center}
		\caption{\label{table1}The Kasner transformations. \textbf{Right Column}. The trajectories described after the action of the Kasner transformations. %They act on the set of the Kasner exponents $a$, $b$, $c$ by suitable permuting them. For the preferred order $a,b,c$, the Kasner transformations have the effect of mapping $xy$ epochs in $ba$ epochs.\newline
\textbf{Central Column} The Kasner transformations $K_i$ acting on the complete billiard domain on the UPHP on the variable $z=u+iv$. They are decomposed in sequneces of simple reflections.
\textbf{Left Column}. The Kasner transformations $K_i$'s evaluated on the absolute at $v=0$.}
\end{table}
%%%%%%%%%%%%%%%%%%%%%%%%%%%%%%%%%%
\section{Implementation of the transfer operators in the unquotiented Billiard system\label{section5a}}
It is possible to analyze the 'Big Billiard' as far as its unquotiented dynamics is concernd, and with its complete phase space, as fron \cite{belinskibook} Chapter 4, without any symmetry-quotienting of the dynamics. It is nevertheless assumed that the geodesic flow analysed is one with geodesics with normalised unit velocity, as the reduction of this degree of freedom of the dynamics descends from the geometricl transformations followed to obtain the UPHP framework.\\
After the implementation of the quantum regime,and, in particular, after the definition of BKL quantum number in the quantum regime, the semiclassical regime can be implemented. The semiclassical Poincar\'e return map is in order to be studied in the unquqotiented big billiard within its complete phase space after the choice of the surface of section corresponding to $v=0$ for the analysis of the energy elvels associated iat the BKL quantum numbers in the semiclassical regime.\\
The BKL quantum numbers in the quantum regime associate to Maass waveforms; in the semiclassical regime, the transfer operator is applied to the BKL-number forms. The energy levels in the semiclassical regime are therefore obtained.

\section{Comparison with the BMG\label{section5}}
%As evident, there are several differences between the dynamics predicted by the action of the BMG, and that described by the SBG.\\
 
%The main difference between the BMG and the SBG is the absence of symmetry-quotienting mechanisms. In fact, the transformation \ref{1z} of the BMG does not coincide with the transformation \ref{R3} of the SBG.
While the BMG is obtained from a desymmetrizzation of the domain of the $SL(2,Z)$ group, the BMG is due to the symmetry-quotienting of the big billiard table according to the presence of the symmetry walls.\\
\\
%As far as the %\textit{dynamical} 
%properties of the small billiard are concerned, it is important to remark that, according to the desymmetrized shape of its domain, no identification is possible between the sides of the small billiard table, such that the different periodic phenomena described by the different $t_{XY^n}$ maps cannot be identified.\\
%On the other hand, as far as the \textit{geometrical} properties of the small billiard are concerned, as already analyzed in \cite{Lecian:2013cxa}, the (several) geometrical transformations that would allow one to recover squared subdomains for the reduced phase space of the SBG do not correspond to any succession of the language code for the SBG.\\  
The BKL map for the small billiard consist of a different number of Weyl reflections, according to the different subregions of the reduced phase space where the first epoch of each small-billiard era is issued from. This property further explains the \textit{physical} content of the $\top_{XY^n}$ maps: as a different succession of matrices is implied for every $\top_{XY^n}$, in particular, those with $n$ odd will contain and odd number of (Weyl) reflections, while those with $n$ even will contain an even number of (Weyl) reflections.\\
As far as the time evolution and cyclic identities are concerned, one remarks that
the Selberg trace formula will not therefore be described according to a suitable conjugacy subclass of the modular group, being
$\Gamma_2(PGL(2,C))$ and $PGL(2,C)$ larger (with respect to the modular group) groups.\\
The cyclic identities of their language codes describing periodic orbits define the parity of the wavefunction solving the WDW equation for this implementation of the Einstein field equations, according to the number of BKL epochs contained in all the BKL eras which the periodic orbit consists of.\\
Taking into account the sequence of simple reflections from Table \ref{table1} allows one to implement the transfer operator on the forms chosen after the BKL quantum numbers with the help of the proper choice of a surface oof section of the Poincar\'e return map on the complete phase space; the Selberg trace formula acts accordingly.
\section{The validation of the trace formulas after the quantum operators\label{secmath}}
The trace formulas are validated after the verification of the existence of transfer operators \cite{ref2}.\\
The topic is developped and applied to the Selberg trace formula in \cite{ref1}. 
The definition of the trace formulas is propedeutically studied after the study of the defintion of th existence of a kernel $K(q, q', E)$ which defines a Fredholm integral equation
\be\label{eqz1}
\psi(q)=\int_{PSS}dq'K(q, q', E)\psi(q')
\ee
which is defined on a one-dimensional Poincar\'e surface of section $PSS$. Poincar\'e surfaces of sections for the dynamics of the flow associated with the desymmetrised $PSL(2, \mathcal{Z})$ domain are studied in \cite{lecian1}.\\
Eq. (\ref{eqz1}) is demonstrated to have non-trivial solution iff $E$ is within the spectrum; it is integrated after the boundary integral method \cite{ref3}.\\
From \cite{ref2}, the semiclassical expression is obtained after substituting the kernel $K$ with the opportune (Bognomolny transfer) operator $T$non the suitably-chosen orbits; the surface of section $\Sigma$ can be chosen after the definitions from \cite{lecian1} (which are suitably for the requests of \cite{belinskibook}, Chapter 4); the trajecotry on which the integrateion id conisdered is defined through the trajectory the action $S_E(q, q')$ follows in the phase space, for which the surface of section is defined \cite{lecian1}, with a total energy $E$, and which connects the point $q$ with the point $q'$.\\
The choice of the one-dimensional surface section in \cite{ref2} is motivated after the analyis of the Markov partition induced after rules of the operators-ordering in the conjugacy classes of the groups considered; is coincides with the definition of the Birkhoff surface of section form \cite{lecian1}, which is motivated after the requirements of the Anosov flow of the system.\\
The semiclassical expression of the Bogomolny transfer operator is here rewritten as
\be
T_E(q, q')=\frac{1}{2i\pi \hbar}\sqrt{\frac{\partial^2 S_e(q, q')}{\partial q, q'}}\cdot e^{\frac{i}{\hbar}S(q, q', E)+i\frac{\pi}{2}\nu}
\ee
The semiclassical Poincar\'e map is defined after the Bogomolny transfer operator ${\bf T}_E(q, q')$ as
\be
{\bf T}_E(q, q')\int_{\Sigma}T_E(q, q')\psi(q)d^Nq
\ee 
being $\Sigma$ the chosen surface of section.\\
The semiclassical limit $\hbar\rightarrow0$is demonstrated to exist iff the spectrum can contain the eigenvalue $1$.\\
It is possible to demonstrate that the operator ${\bf T}$ admits an invariant function $\tilde{\psi}$. The invariant function $\tilde{\psi}$ is written as
\be
\tilde{\psi}(q')\equiv\int_{\Sigma}T_e(q', q)\tilde{\psi}(q)d^Nq.
\ee
The invariant function $\tilde{\psi}$ is defined after the compatibility condition of the considered Fredholm determinant as
\be\label{z2}
det(\hat{1}-{\bf T}_E)=0.
\ee
The solutions of Eq. (\ref{z2}) are proven to coincide with the dynamical zeta function as  infinite product over all the periodic
orbits\cite{ref6}, \cite{ref7} under the suitable conditions \cite{ref8}, \cite{ref9}, \cite{ref10}, \cite{ref11}.\\
The quantum maps ${\bf T}_s$ is defined after the definition of its adjoint operator ${\bf U}$, after the requirements of \cite{ref4}.\\
In \cite{ref1}, the existence of an ergotic measure on the Banach spaces on which these operators acts is ensured.\\
The qualities of the defined operators to be applied on irrational are discussed still in \cite{ref4}; furthermore, in \cite{ref4}, selected topics about the meromorphic continuation of the objects are calculated.\\

As far as the modular domain is concerned \cite{reff7}, since ${\bf T}_s$ is the semiclassical operator, the Selberg zeta function $\zeta_s$ can be written as a function of generalised Fredholm determinant, which can be analogous to Eq. (\ref{z2}), i.e.
\begin{equation}
\zeta_s=det(\hat{1}-{\bf T}_s)det(\hat{1}+{\bf T}_s)
\end{equation}
where energy $E$ is parameterised as for the Maass eigenforms $E=s(s+1)$.\\
From \cite{reff9}, the Selberg zeta function can be written as a function of the generalisd determinant
\be
\zeta_s=\theta(s)det(\hat{1}-{\bf T}_s),
\ee
being $\theta_s$ a function determined after the properties of particular orbits.\\
The derivation of the equivalence between the function determinant and the Selberg zeta function is valid in the absence of singular orbits in the pertinent expansion of the zeta function.\\
As far as the billiard description of the geodesics flow is concerned, the presence of particular periodic configurations should not be overlooked, such as the periodic orbits consisting of the trajectory following the sides of the billiard table: these last types of trajectories as excluded form thee analysis after the definition of \cite{corn1982}, in which the corner points of the billiard are excluded form the definition of trajectories ab initio. 
The particular case of the Artin billiard, which corresponds to the BKL small billiard, is discussed still in \cite{ref2}.\\
The topics here proven are further discussed in \cite{ref5}.\\
The (Banach) space of observables for the semcilassical descriptions are discussed in \cite{ref5}, \cite{reff9}, \cite{reff10}.\\
The description of the quantum properties of the Maass wavefunctions (on which the operators acts) is derived form the Hecke theory in \cite{dok1} and proven to descend from the properties of automorphisms on the trees.

\section{Quantum regime\label{section6}}
The semiclassical Poincar\'e return map \cite{bogomolny1}, %\cite{ruelle}, 
is defined as
\begin{equation}\label{poinc}
\tau_E\psi(q')=\int_{\Sigma}\tau_E(q',q)\psi(q')d^Nq
\end{equation}
where the integral is performed on a surface of section $\Sigma$, and is extended to the corresponding degrees of freedom of
the phase space,
with $\tau_E$ a billiard map obeying the consistency equation
\begin{equation}\nonumber
0=(1+\tau_E)(1-\tau_E)
\end{equation}
and defines the Selberg $\zeta$ function according to the generalized \cite{gr1956} determinants $det(1-\tau_E)$ and $det(1-\tau_E)$ as
\begin{equation}\label{maass1}
\zeta(s)=det(1+\tau_E)det(1-\tau_E)=\theta(s)det(1-\tau_E)
\end{equation}
The expression of (\ref{poinc}) for the case of cosmological billiards is then expressed according to the angle $\theta$ defined in Figure \ref{figuraustar} as \cite{lecian2013qw} , or by its expression as a function of $u^*$, i.e. the value $u=const$ that defines a generic Poincar\'e section different from the billiard sides, and which parameterized the energy-shell reduced Liouville measure.\\
The classification of all the periodic orbits according to the different maps allows one to reconstruct the complete spectrum of the operator $\tau_E$, which is the same as its self-adjoint operator $U_E$, which acts on the Maass waveforms.\\
\vspace{9mm}
The density levels $dE$ of the quantum systems and their classical periodic orbits are related at the semiclassical transition by the Selberg trace formula
\begin{equation}
dE=\bar{dE}+dE^{osc},
\end{equation}
\begin{equation}
dE^{osc}=\sum_nA_n(E)e^{\tfrac{iS_n}{2\pi h}},
\end{equation}
being  $\bar{dE}$ and $dE^{osc}$ the average (i.e. on containing singular orbits) contribution and the one corresponding to periodic orbits, respectively, with $S_n$ expansion of the action and $A_n$ the corresponding coefficient within hte sum on perioidic orbits $n$. The number of epochs in each BKL eras constituting periodic orbits for cosmological billiards define the parity (the sign) of the eigenvalues for the quantum-mechanical description of cosmological billairds on the UPHP, corresponding to the BKL (towards the cosmological singularity) asymptotic limit of the WDW equation, solved by the (Maass) wavefunctions for cosmological billiards \cite{Lecian:2013cxa}.\\

\section{Outlook and perspectives\label{section7}}

\paragraph{Outlook}The mathematical analysis of the discretized dynamics of Cosmological Billiards has to be considered as relevant for the physical characterization of the quantum regime, the semiclassical transition and the classicalized states of a generic Universe with respect to the present observed Universe.\\
The definition of such characterization is needed for the comparison of the experimental evidence providing definition about the evolution of the Universe and the external (i.e. 'on the r.h.s. of the E.f.e.'s) which have to be supposed to have taken place as modifying the oscillatory behavior of a Generic Universe with respect to the present observed values of anisotropy, as well as for the anisotropy rates of the statistical distribution of matter densities as far as the investigation on Astrophysical scales is concerned: the introduction of isotropization mechanisms of chaotic models, and those of quasi-isotropization \cite{9}, can allow for a comparison with observational evidence, as well as the hypothesis of some inflation-generating mechanism \cite{9a}.\\
The feaures of the spectrum of the energy levels of the wavefunction allow one to test the phenomenological effects obtained in Quantum-Gravity models \cite{8}, as far as the possible deformations of the background (geometrical) space are concerned, and allow one to compare the effectiveness of such Quantum-Gravity motivated investigations in modifying the chaotic properties of the billiard systems with the effects of classical scalar fields and vector fields \cite{8a}.
The introduction of isotropization mechanisms of chaotic models, and those of quasi-isotropization \cite{9}, can allow for a comparison with observational evidence, as well as the hypothesis of some inflation-generating mechanism \cite{9a}.\\
%A complementary investigation line is constituted by \cite{mamar14} and \cite{mamar15}.

\paragraph{Perspectives}The analysis of the solution of the iterations of Cosmological Billiard Maps which define periodic orbits for cosmological billiards is fundamental for the comparison of the known results for the mathematical properties of periodic trajectories on domains on curved hyperbolic spaces, whose dynamics does not result as a symmetry-quotienting of a pre-existing dynamical system.\\
The implementation of quantum statistical maps for Cosmological Billiards is based on the comparison between the geometrical properties of these systems, which provide the analysis with a suitable group-theoretical structure, and the symmetries of the dynamics, for which the symmetries of the metric tensor define a 'smaller' class of transformations, which characterizes the statistical description. This method of investigation, the Jacquet-Langlands correspondence, is framed in the broader Langland programme \cite {4b}.\\
Indeed, for cosmological billiards in $4=3+1$ spacetime dimensions, the the semiclassical Poincar\'e return map defined by Eq. (\ref{poinc}) is defined by a map $\tau_E$ for a given energy level corresponding to the classical configuration of energy $E$ (at which the classical reduced phase space corresponds).
More in detail, the operator $\tau_E$ corresponds to any of the classical billiard map defined for cosmological billiard, i.e. either the big billiard unquotiented map, or the big billiard Kasner quatiented maps, such as the BKL epoch map, the BKL era-transition map and the CB-LKSKS map, or the small billiard unquotiented map, or the small billiard BKL map.\\
The expression of (\ref{poinc}) for the case of cosmological billiards is then expressed according to the angle $\theta$ defined in Figure \ref{figuraustar} that defines a generic Poincar\'e section different from the billiard sides.%, and that parameterizes the energy-shell reduced Liouville measure.\\
The classification of all the periodic orbits according to the different maps allows one to reconstruct the complete spectrum of the operator $\tau_E$.\\
Buondary conditions for cosmological billiards have already been thoroughly discussed in the literature.
Both Neumann and Dirichlet boundary conditions have been proposed and motivated, according to different features that had to be described.\\
From (\ref{maass1}), one learns that the two different conditions, i.e. $(1+\tau_E)$ or $(1-\tau_E)$ correspond to Neumann boundary conditions and to Dirichlet boundary conditions, which correspond, on their turn, to odd wavefunctions or to even wavefunctions.\\
\\ 
It is crucial to remark that the identification of Eq. (\ref{poinc}) to a semiclassical version of the BKL map operators for a fixed energy shell, and the identification of Eq. (\ref{maass1}) with the choice of boundary conditions is restricted to  either the surface of section $\Sigma$ corresponds to one side of the billiard (for the purposes of this analysis, the side $b$ of the big billiard), or the surface of section $\Sigma$ does not correspond to a side of the billiard table, but the knowledge of both the unquotiented dynamics and the         requested maps allows one to recast the proper geodesic flow through $\Sigma$, i.e. the one corresponding to that bouncing onto a side. 
The classical description of the cosmological billairds on the UPHP ant with its restricted phase space is obtained by fixing a given energy at which the Hamiltonian flow is calculated.\\
The operator $\tau_E$ defined in Eq. (\ref{poinc}) can therefore be interpreted as the operator that, for each classical energy level $E$, acts on a semiclassical wavefunction (semiclassical in the sense that it is evaluated on a classical BKL configuration corresponding to a geiven sequence of epochs and eras).\\
At each energy level $E$, the operator(s) $\tau_E$ leave invariant (except for a $\pm$ sign) the eigenfunctions of the quantum eigenvalue problem; the Selberg eigenfunctions are defined as generalizing the Riemann $\zeta$ function to closed (i.e. periodic) orbits.% (instead of prime numbers).\\
The BKL map operators define the complete set of periodic orbits of cosmological billiards (from a classical point of view).\\
The corresponding system of operators $\tau_E$ can therefore be interpreted as the semiclassical operator which extracts an eigenvalue for each closed geodesics, which correspond to periodic orbit, according to it content of BKL epochs, BKL eras and the chosen symmetry-quotienting mechanism, and whose complete spectra are now classified.
\\

\section{Concluding remarks\label{section8}}
The aim of this investiagtion has been to define periodic phenomena for Cosmologial Billiards in $4=3+1$ spacetime dimensions. In particular, quantum BKL maps and the semiclassical BKL maps have eigenvalues which define the sign acquired by the wavefunctions for cosmological billairds, whose parity is defined by the corresponding periodicity phenomena.\\ 
The paper is organized as follows.\\
In the Introduction (\ref{section1}), Cosmological Billiards are introduced.\\
In Section \ref{section2}, the Billiard Modular Group is defined.\\
The Big Billiard Group is defined in Section \ref{section3}, %and the Small Billiard in Section \ref{section4},
 for which the differences with the Billiard Modular Group are outlined in Section \ref{section5} as those characterizing the symmetries of the solutions to the Einstein field equations.\\
The mathematical tools propedeutical of the study of the Selberg trace formula of cosmological billiards are gathered in Section \ref{secmath}:
For the cosmological billiards, the Selberg trace formula is assessed in Section \ref{section6}: for the pertinent billiard groups, the conjugacy subclasses are classified, and periodic orbits are therefore spelled out and analyzed.\\
The sign acquired by the Maass wavefunctions under the quantum BKL maps %\ref{ruelle2} 
characterizes the Selberg trace formula for Cosmological Billiards, which also defines the energy-level densities according to the quantum BKL maps for periodic orbits.\\ %As analyzed in Section \ref{section6} and initially commented in Section \ref{section7},
The  quantum BKL map operators are the same self-adjoint operator of the quantum BKL operators defined on the solution of the minisuperspace reduction of the WDW equation for cosmological billiards. This result strictly applies for the analysis of %\cite{4b} and \cite{gr1956} and further specifies the analysis of %\cite{ruelle2} and 
\cite{bogomolny1} for Cosmological Billiards.

%%%%%%%%%%%%%%%%%%%%%%%%%%%%%%%%%%%%%%%%%%%%%%%%%%%%%%%%%%%%%%%%%%%%%%%%%%%%%%%%%%   

%%%%%%%%%%%%%%%%%%%%%%%%%%%%%%%%%%

\section*{Acknowledgments}
OML would like to thank Prof. H. Nicolai for having suggested Ref. \cite{bogomolny2}, and for having warmly encouraged the classification of periodic phenomena for cosmological billiards in $4=3+1$ dimensions.\\
The work of OML was partially supported by the research grant 'Reflections on the Hyperbolic Plane' from the
Albert Einstein Institute.\\
OML acknowledges the Albert Einstein Institute- MPI warmest hospitality during the corresponding stages of this
work.% A. Kleinschmidt is kindly tanked for a strictest discussion of the hyperbolicity of the classical billiard maps is Sections \ref{section2}, \ref{section3}, \ref{section4}.\\
%OML is grateful to Prof. Y.I. Manin for having brought to the attention Ref. \cite{connes}.

\appendix
\section{The small-billiard map\label{appendixa}}
The CB-LKSKS map for the small billiard, $t_{CB-LKSKS}$, named after Chernoff- Barrow- Lifshitz- Khalatnikov- Sinai- Khanin- Shchur, is defined by two different kinds of transformations, i.e., 
\begin{subequations}\label{tsb}
\begin{align}
&t^{1,2}z=T^{-1}SR_1T^{-n+1}z,\ \ {\rm for} (u^+, u^-)\in S^{1}_{ba} {\rm and} (u^+, u^-)\in S^{2}_{ba},\label{tsb1}\\
&t^{2',3,3'}z=T^{-1}SR_1T^{-n+1}R_3z,\ \ {\rm for} (u^+, u^-)\in S^{2'}_{ba}, (u^+, u^-)\in S^{3}_{ba}, {\rm and} (u^+, u^-)\in S^{3'}_{ba}\label{tsb2}.
\end{align}
\end{subequations}
which act on the subregions of the reduced phase space $S^1_{ba}$, $S^2_{ba}$, $S^3_{ba}$, $S^{2'}_{ba}$ and $S^{3'}_{ba}$ defined as
\begin{itemize}
     \item $S^1_{ba}:\ \ u^-<-\Phi,\ \ u^+>u_\alpha(u^+),\ \ -\Phi<u^-<-1,\ \ u^+>u_\gamma(u^+)$;
     \item $S^2_{ba}:\ \ -\Phi<u^-<-1,\ \ u_\alpha(u^+)<u^+<u_\gamma(u^+)$;
     \item $S^{2'}_{ba}:\ \ u^-<-2,\ \ 0<u^+<u_\alpha(u^+),\ \ -2<u^-<-\Phi,\ \ u_\gamma(u^+)<u^+<u_\alpha(u^+)$;
     \item $S^{3}_{ba}:\ \ -2<u^-<-\Phi,\ \ u_\gamma(u^+)<u^+<u_\beta(u^+),\ \ -\Phi<u^-<-1,\ \ u_\alpha(u^+)<u^+<u_\beta(u^+)$; \\
     \item $S^{3'}_{ba}:\ \-2<u^-<-1,\ \ 0<u^+<u_\beta(u^+)$;
\end{itemize}
where the functions
\begin{itemize}
     \item $u_\alpha(u^+):\ \ u^+=-\tfrac{1}{u^-}$;
     \item $u_\beta(u^+):\ \ u^+=-\tfrac{u^- + 2}{2u^- + 1}$; 
     \item $u_\gamma(u^+):\ \ u^+=-\tfrac{u^-+2}{u^-+1}$;
\end{itemize}
are defined. In particular, the function $u_{\gamma}(u^+)$ corresponds to the image of the function $u_\alpha(u^+)$ according to the transformation $B$.    
\subsection{The language code for the small billiard group}
The following language code for the small billiard is obtained
\begin{enumerate}
	\item $R$, $B$, $G$, $BR$, $BG$, $RG$, $RB$, $GR$, $GB$;
	\item $RBR$, $BGR$, $RGB$;
	\item $\top_{XY^n}$.
\end{enumerate}
As one can straightforward verify, the sequence $RGR$ is not allowed.\\
The trajectories $\top_{XY^n}$ are a succession of epochs such that the first epoch starts form a $R$ wall, and the last epoch ends on a $R$ wall; between these two epochs, bounces between the $G$ and the $B$ wall take place, such that $n$ trajectories are present.\\
As classified in \cite{}, the reduced phase space for the small billiard table is characterized curvilinear domains. In particular, is is possible to further divide these domain according to the preimage of the transformations (\ref{t}) on the $RG$ subdomain and on the $RB$ with respect to the $BR$ subdomain and to the $GR$ one.\\
This way, the trajectories $\top_{XY^n}$ are defined as a succession of transformations of the kind $\top_{XY^n}\equiv RXY...X'Y'$, where the transformations $X$, $Y$, $X'$, $Y'$ can be $B$ or $G$. More in detail, they are defined on the small billiard reduced phase space as
\begin{enumerate}
	\item $\top_{RG^n}\equiv RGB...BG$, $n$ odd, $u^+>1, u^-<-1/2, (u_\alpha(u^+), u_\beta(u^+), u_b^n(u^+), u_a^n(u^+))$;
	\item $\top_{RG^n}\equiv RGB...GB$, $n$ even, $u^+>1, u^-<-1/2, (u_\alpha(u^+), u_\beta(u^+), u_b^{n-1}(u^+), u_a^{n+1})(u^+)$; 
	\item $\top_{RB^n}\equiv RBG...GB$, $n$ odd, $u^+<0, u^->-1/2, (u_\alpha(u^+), u_\beta(u^+), U_b^n(u^+), U_a^n(u^+))$;
	\item $\top_{RB^n}\equiv RBG...BG$, $n$ even, $u^+<0, u^->-1/2, (u_\alpha(u^+), u_\beta(u^+), U_b^{n-1}(u^+), U_a^{n+1})(u^+)$;
\end{enumerate}
The functions $u_b^n(u^+)$, $u^a_n(u^+)$, $U^b_n(u^+)$, $U_a^n(u^+)$ are defined as
\begin{itemize}
	\item $u_b^n(u^+):\ \ u^+=\tfrac{1}{2}\frac{-2nu^-+2u^-+n^2-2n+5}{-2u^-+n-1}$;
	\item $u^a_n(u^+):\ \ u^+=\tfrac{1}{2}\frac{-4n+7-2nu^-+4u^-+n^2}{-2-2u^-+n}$;
	\item $U^b_n(u^+):\ \ u^+=-\tfrac{1}{2}\frac{2nu^--2u^-+n^2-2n+5}{2u^-+n-1}$;
	\item $U_a^n(u^+):\ \ u^+=-\tfrac{1}{2}\frac{3+2nu^-+2n^2}{2u^-+n}$,
\end{itemize}
and correspond to the preimage of the $RB$ and $RG$ regions of the reduced phase space according to the pertinent combination of transformation $G$ and $B$.

\subsection{Action of the SBG on the oriented endpoints.}
Periodic orbits for the SBG are defined according by imposing that the oriented endpoints obey the condition
\begin{equation}
\top_{XY^n}u\equiv u.
\end{equation}
According to the classification of the sequences $\top_{XY^n}$, the following quadratic equations with integer coefficients are obtained for the endpoints, respectively:
\begin{enumerate}
	\item $\top_{RG^n}$, $n$ odd, $u^2-mu+1=0$, with $m\equiv \tfrac{n}{2}-1$;
	\item $\top_{RG^n}$, $n$ even, $u^2-mu+1=0$, with $m\equiv \tfrac{n}{2}$; 
	\item $\top_{RB^n}$, $n$ odd, $u^2+mu+1=0$, with $m\equiv \tfrac{n+1}{2}$;
	\item $\top_{RB^n}$, $n$ even, $u^2+mu-1=0$, with $m\equiv \tfrac{n}{2}$.
\end{enumerate}
These transformations are always hyperbolic, since their discriminants $\Delta_m$ reads
\begin{enumerate}
	\item $\Delta_m\equiv\sqrt{m^2+4}$ for $t_{RG^n}$ and $t_{RB^n}$, with $n$ even;
	\item $\Delta_m\equiv\sqrt{m^2-4}$ for $t_{RG^n}$ and $t_{RB^n}$, with $n$ odd, 
\end{enumerate}
as the minimum number of epochs in each $\top_{XY^n}$ era is $n\le 3$.\\
The periodic orbits defined by an even number $n$ of epochs allow for a continued-fraction decomposition analogous to that obtained in the case of the Golden Ratio and of the 'silver ratios' for the big billiard. In the case of the big billiard, the phase-space points that define periodic orbits identified by these 'ratios' are placed along the function $u_\gamma(u^+)$ of the starting box $F_{ba}$.\\
In the case of an odd number of epochs, this decomposition does not hold any more. Furthermore, it is not possible to define any transformation able to map these trajectories to this kind of decomposition.\\
\paragraph{The epoch map for the unquotiented small billiard}
Collecting all the ingredients together, it is possible to generalize the content of the sequences $\top_{XY^n}$ and to establish a map for the unquotiented variable $u\equiv u^\pm$ relating each first epoch of the small-billiard eras to each last epoch of the small-billiard eras, denoted by the phase-space points $u_F\equiv u^\pm_F$ and $u_L\equiv u^\pm_L$, respectively, as
\begin{equation}\label{ll1}
u_L=\top_{GY^n}u_F\equiv(-1)^{n}(u-m),
\end{equation}
and
\begin{equation}\label{ll2}
u_L=\top_{BY^n}u_F\equiv(-1)^{n}(u+m),
\end{equation}
with $m$ defined in the above.\\
The era transition map is obtained by composing the epoch map with the transformation $S$, which accounts for the bounce of the $R$, side, such that the first epoch of the successive era is defined by the phase space points of the reduced phase space $u'\equiv u^{\pm'}$, i.e.
\begin{equation}
u'\equiv S\top_{XY^n}.
\end{equation}

\end{document}